\documentclass[english,aps,pra,reprint,superscriptaddress]{revtex4-1}
\usepackage[T1]{fontenc}
\usepackage[latin9]{inputenc}
\usepackage{geometry}
\usepackage{color}
\usepackage{babel}
\usepackage{float}
\usepackage{amsmath}
\usepackage{amssymb}
\usepackage{graphicx}
\usepackage{wasysym}
\usepackage[unicode=true,pdfusetitle,
 bookmarks=true,bookmarksnumbered=false,bookmarksopen=false,
 breaklinks=false,pdfborder={0 0 0},pdfborderstyle={},backref=false,colorlinks=true]
 {hyperref}
\hypersetup{
 linkcolor=blue, urlcolor=blue, citecolor=blue}

\makeatletter
\usepackage{braket}
\usepackage{dsfont}
\usepackage{bm}
\date{}

\makeatother

\begin{document}
\title{Employing Non-Markovian effects to improve the performance of a quantum
Otto refrigerator}
\author{Patrice A. Camati}
\email{ patrice.camati@neel.cnrs.fr}

\affiliation{Université Grenoble Alpes, CNRS, Grenoble INP, Institut Néel, 38000
Grenoble, France}
\author{Jonas F. G. Santos}
\email{jonas.floriano@ufabc.edu.br}

\affiliation{Centro de Ciências Naturais e Humanas, Universidade Federal do ABC,
Avenida dos Estados 5001, 09210-580 Santo André, São Paulo, Brazil}
\author{Roberto M. Serra}
\email{serra@ufabc.edu.br}

\affiliation{Centro de Ciências Naturais e Humanas, Universidade Federal do ABC,
Avenida dos Estados 5001, 09210-580 Santo André, São Paulo, Brazil}
\begin{abstract}
The extension of quantum thermodynamics to situations that go beyond
standard thermodynamic settings comprises an important and interesting
aspect of its development. One such situation is the analysis of
the thermodynamic consequences of structured environments that induce
a non-Markovian dynamics. We study a quantum Otto refrigerator where the 
standard Markovian cold reservoir is replaced by a specific engineered
cold reservoir which may induce a Markovian or non-Markovian dynamics
on the quantum refrigerant system. The two dynamical regimes can be
interchanged by varying the coupling between the refrigerant and the
reservoir. An increase of non-Markovian effects will be related to
an increase of the coupling strength, which in turn will make the
energy stored in the interaction Hamiltonian, the interaction energy,
increasingly relevant. We show how the figures of merit, the coefficient
of performance and the cooling power, change for non-negligible interaction
energies, discussing how neglecting this effect would lead to an overestimation
of the refrigerator performance. Finally, we also consider a numerical
simulation of a spin quantum refrigerator with experimentally feasible
parameters to better illustrate the non-Markovian effects induced
by the engineered cold reservoir. We argue that a moderate non-Markovian
dynamics performs better than either a Markovian or a strong non-Markovian
regime of operation.
\end{abstract}
\maketitle

\section{Introduction}

The theoretical description of thermal machines was fundamental to
the development of classical thermodynamics, providing an operational
understanding of the second law as established by Thomson
(Lord Kelvin)~\citep{Thomson1851} and Carnot~\citep{Carnot}.
Moreover, the thermodynamic characterization of heat engines and refrigerators
is essential to engineering since it provides tools for estimating
the performance of such machines~\citep{=0000C7engelbook2015,Kondepudibook2015}.
In the same perspective, it is expected that the development of quantum
thermodynamics will play a similar role in quantum engineering to
the development of quantum technologies~\citep{Dowling2003,Georgescu2012,Jones2013}.

Quantum thermal machines are excellent platforms to test results from
quantum thermodynamics~\citep{Esposito2009,Campisi2011,Kosloff2013,GelbwaserKlimovsky2015,Millen2016,Vinjanampathya2016,Ribeiro2016},
transforming heat into work and vice versa. In the quantum heat engine
configuration, the purpose is to extract the largest amount of work
by absorbing the least amount of heat from a hot source. On the other
hand, in the quantum refrigerator setting, the goal is to absorb the
largest amount of heat from a cold source by injecting the least amount
of work into the refrigerant. The performance of the former is characterized
by the thermodynamic efficiency and the power output, while that of
the latter is characterized by the coefficient of performance (COP) and the cooling
power. The theoretical underpinnings of the description of quantum
thermal machines dates back to the late 1950s with the early works
of Scovil and Schulz-DuBois~\citep{Scovil1959,Geusic1967}. Since
then, several theoretical investigations on quantum heat engines and
refrigerators have been carried out~\citep{Alicki1979,Kosloff1984,Geva1992,Geva1992b,Geva1996,Kosloff2000,Scully2002, S=0000E1nchezSalas2004,Feldmann2006,Ting2006,Quan2007,Allahverdyan2008,Linden2010,Brunner2012,
Zhang2018, Obinna2016,Erdman2019A,Brunner2015,Kosloff2013-1,Wookds2015,Kosloff2012,Camati2019}.

From the experimental point of view, two microscopic classical heat
engines~\citep{Ro=0000DFnagel2016,Volpe2017}, a quantum refrigerator
with trapped ions~\citep{Maslennikov2017} and a spin quantum engine
in nuclear magnetic resonance (NMR)~\citep{Peterson2018}, have been
recently implemented. Additionally, a quantum Otto cycle on a nano-beam
working medium~\citep{Klaers2017} and a quantum heat engine using
an ensemble of nitrogen-vacancy centers in diamonds~\citep{Klatzow2018}
have been reported. In particular, the experiment reported in Ref.~\citep{Klaers2017}
employs a squeezed thermal reservoir and demonstrated that the efficiency
of the quantum heat engine (that explores squeezing) may go beyond
the standard Carnot efficiency, corroborating theoretical expectations~\citep{Huang2012,Abah2014,Correa2014,Ro=0000DFnagel2014,Alicki2014,Zhang2014-1,Alicki2015,Long2015,Niedenzu2016,Manzano2016,Kosloff2017,Latune2018}.

Interaction with a squeezed thermal reservoir is not the only generalized
process that goes beyond the typical settings in classical thermodynamics.
For instance, engineered reservoirs~\citep{Abah2014}, such as including
coherence~\citep{Scully2003}, have also been addressed, evidencing
that quantum properties can be used to enhance the performance of
quantum heat engines and refrigerators when compared to their conventional
counterparts.

Reservoir engineering may be useful and important in different physical
setups, for instance, for cooling phonons~\citep{Lemond} and in
circuit quantum electrodynamics~\citep{Gross01}. Quantum fluctuation
theorems can be probed by using engineered reservoirs~\citep{Carvalho}.
In the context of quantum thermodynamical processes an approach for
reservoir engineering inducing non-Markovian dynamics has been considered
in Ref.~\citep{Raja2017}, where the complete reservoir structure
is composed of a Markovian part plus a two-level system.

In recent years, advances in quantum technologies and quantum control
allowed the study of effects beyond the Born-Markov approximation
in open systems, enabling tests of memory effects in decoherence dynamics~\citep{Liu001}.
Although, in general, the non-Markovian aspects of the dynamics are
associated with a strong coupling between system and reservoir, a
non-Markovian dynamics may be observed in a weak-coupling regime,
for instance, when the reservoir has a finite size (structured reservoir)~\citep{Sampaio}.
Recently, non-Markovian aspects in quantum thermodynamics have been
studied from different points of view, for instance, in the context
of thermodynamic laws and fluctuations theorems~\citep{Whitney2017},
in non-equilibrium dynamics~\citep{Chen2017}, and their effects
on the entropy production of non-equilibrium protocols~\citep{Bhattacharya2017,Marcantoni2017,Popovic2018}.
From the perspective of quantum thermal machines, a promising avenue
is emerging with recent theoretical results illustrating how to use
memory effects to improve the performance on quantum thermodynamic
cycles~\citep{Thomas2018,Abiuso2019,Paternostro2019}.

In this paper, we are interested in studying and quantifying the performance
of a quantum Otto refrigerator in which the particular structure of
the engineered cold reservoir may generate a non-Markovian dynamics
on the quantum refrigerant. Due to the nature of our engineered cold
reservoir, the interaction energy between the refrigerant and the
reservoir is not negligible, having an important role in the performance.
Such an impact has been previously addressed for quantum heat engines
where the interaction energy was considered as an additional cost
for the performance~\citep{Newman2017,Newman2019}. We show how the
expressions for the COP and cooling power change, including this contribution,
and discuss how the interaction energy impacts the performance of
the refrigerator. Employing incomplete thermalization with the engineered
cold reservoir (at the finite-time regime), we show that memory effects
(non-Markovianity) serve as a resource to increase the performance
of the quantum Otto refrigerator, provided we have a sufficiently
high control of the parameters involved in the cycle, for instance,
the time allocation in each stroke. We show that the performance does
not always improve as one increases the memory effects but an intermediate
no-Markovian dynamics corresponds to the best performance. In order
to illustrate our results, we consider a numerical simulation of a
single-qubit quantum Otto refrigerator model.

This paper is organized as follows. In section~\ref{sec:Model-of-non-Markovianity}
we discuss the model of an engineered cold reservoir employed in the
quantum Otto refrigerator. Section~\ref{sec:Injected-power-and}
is devoted to present and discuss the performance of the refrigerator,
i.e., the role of memory effects in the COP, cooling power, and injected
power. We provide analytical and numerical results employing experimentally
feasible parameters to illustrate that memory effects may improve
the performance of a quantum refrigerator. Finally, in the section
\ref{sec:Conclusions} we draw our conclusions and final remarks.

\section{Quantum Otto refrigerator with structured cold reservoir\label{sec:Model-of-non-Markovianity}}

\subsection{Model of the structured cold reservoir\label{subsec:Model-of-structured}}

\begin{figure}[t]
\begin{centering}
\includegraphics[scale=0.65]{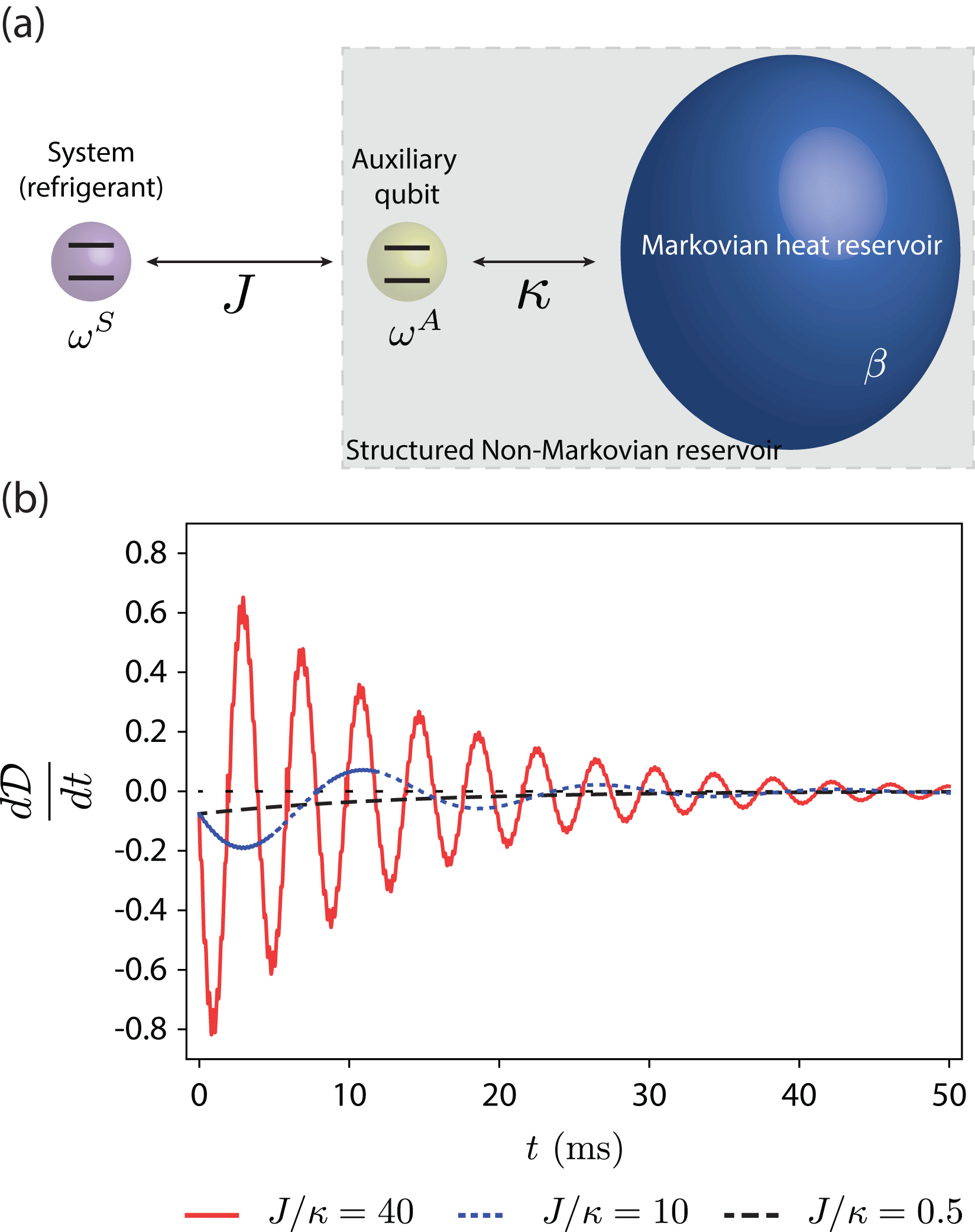}
\par\end{centering}
\caption{(a) Illustration of the structured cold heat reservoir which can induce
a non-Markovian dynamics on the refrigerant of the quantum Otto refrigerator.
It is composed of two parts, a Markovian heat reservoir and a two-level
system (auxiliary qubit). (b)Time derivative of the trace
distance between {the $\left|0\right\rangle $ and
$\left|1\right\rangle $ initial states for the system} for different
values of the ratio $J/\kappa$, 40 (red {solid} line),
10 (blue {dotted} line), and 0.5 (black {dashed}
line) in order to indicate the non-Markovian and Markovian regimes
of the model.}
 \label{Fig01}
\end{figure}

Before describing the quantum Otto refrigerator, we detail the model
of the engineered cold reservoir, which induces a non-Markovian dynamics
on the refrigerant substance. The cold reservoir is composed of two
parts, a Markovian heat reservoir and a two-level system (henceforth
referred as auxiliary qubit), as depicted in Fig.~\ref{Fig01}(a). The
system (quantum refrigerant) will be regarded as a qubit, which interacts
with the cold reservoir by means of the coupling with the auxiliary
qubit. The Hamiltonian of the full composite system is given by

\begin{equation}
H^{SAM}=H^{SA}+\sum_{i}\hbar\omega_{i}b_{i}^{\dagger}b_{i}+\hbar\kappa\sum_{i}g_{i}\left(\sigma_{+}^{A}b_{i}+\sigma_{-}^{A}b_{i}^{\dagger}\right),\label{eq:total hamiltonian}
\end{equation}
where $S$, $A$, and $M$ stand for system, auxiliary qubit, and
Markovian reservoir, respectively; $b_{i}$ ($b_{i}^{\dagger}$) is
the bosonic annihilation (creation) operator of the $i$th oscillator,
which satisfies $\left[b_{i},b_{j}^{\dagger}\right]=\delta_{ij}$;
$\kappa g_{i}$ are the coupling constant; $\omega_{i}$ are the
frequencies; and $\sigma_{\pm}=\left(\sigma_{x}\pm i\sigma_{y}\right)/\sqrt{2}$,
where $\sigma_{x,y,z}$ are the Pauli matrices. The coupling between
the auxiliary qubit and the reservoir is assumed to be weak and flat
so that the dynamics is Markovian. The two-qubit Hamiltonian is given
by 
\begin{equation}
H^{SA}=\sum_{i=S,A}\frac{\hbar\omega^{i}}{2}\sigma_{z}^{i}+\hbar J\sigma_{x}^{S}\sigma_{x}^{A},\label{eq:two qubit hamiltonian}
\end{equation}
where $\hbar\omega^{S}$ and $\hbar\omega^{A}$ are the energy gaps
of the system $S$ and auxiliary qubit $A$, respectively, and $J$
is their coupling constant. Depending on the relative value between
$J$ and $\kappa$, the dynamics of the refrigerant may become non-Markovian~\citep{Raja2017}.
For further reference we denote by $H_{\text{int}}^{SA}$ the last
term on the right-hand side of Eq.~(\ref{eq:two qubit hamiltonian}),
i.e., the interaction Hamiltonian between the refrigerant and the
auxiliary qubit.

When the bosonic modes are in the thermal state it is possible to
show that the two-qubit system $SA$ evolves under the Lindblad master
equation~\citep{Raja2017}
\begin{align}
\frac{d}{dt}\rho_{t}^{SA}= & -\frac{i}{\hbar}\left[H^{SA},\rho_{t}^{SA}\right]+\nonumber \\
 & +\sum_{i=1,2}\gamma^{\downarrow}\left(\epsilon_{i}\right)\left(L_{i}\rho_{t}L_{i}^{\dagger}-\frac{1}{2}\left\{ L_{i}^{\dagger}L_{i},\rho_{t}\right\} \right)\nonumber \\
 & +\sum_{i=1,2}\gamma^{\uparrow}\left(\epsilon_{i}\right)\left(L_{i}^{\dagger}\rho_{t}L_{i}-\frac{1}{2}\left\{ L_{i}L_{i}^{\dagger},\rho_{t}\right\} \right),\label{eq:master equation}
\end{align}
where $\gamma^{\downarrow}\left(\epsilon_{i}\right)=\frac{\pi}{2}\mathcal{J}\left(\epsilon_{i}\right)\left[1+n_{\text{BE}}\left(\epsilon_{i}\right)\right]$
and $\gamma^{\uparrow}\left(\epsilon_{i}\right)=\frac{\pi}{2}\mathcal{J}\left(\epsilon_{i}\right)n_{\text{BE}}\left(\epsilon_{i}\right)$
are the decay rates, with spectral density $\mathcal{J}\left(\omega\right)=\kappa/\pi$
and Bose-Einstein distribution $n_{\text{BE}}\left(\omega\right)=\left(e^{\beta\hbar\omega}-1\right)^{-1}$,
and $L_{i}$ are the Lindblad operators (see~\ref{sec:Master-Equation-of}
for more details). The quantities $\hbar\epsilon_{1}$ and $\hbar\epsilon_{2}$
are the energy gaps between different energy levels of the Hamiltonian
$H^{SA}$ (see Fig.~\ref{fig:Transition-frequencies}).

There are different notions of non-Markovian dynamics for quantum
processes~\citep{Wolf2008,Breuer2009,Laine2010}. Independent of
the precise definition employed, non-Markovianity is always related
to the concept of memory in the dynamics. Here, we will adopt the
following notion: a system undergoes a non-Markovian dynamics if there
is a flow of information from the environment to the system. In particular,
we consider the trace distance as a measure of distinguishability
(information)~\citep{Breuer2009}. For two arbitrary states $\rho$
and $\sigma$, their trace distance is given by~\citep{ChuangBook}
\begin{equation}
\mathcal{D}\left(\rho,\sigma\right)=\frac{1}{2}\text{Tr}\left[\sqrt{\left(\rho-\sigma\right)^{\dagger}\left(\rho-\sigma\right)}\right].
\end{equation}

We denote by $\Lambda_{t,0}\left[\rho\left(0\right)\right]=\rho\left(t\right)$
the one-parameter family of dynamical maps. The dynamics is non-Markovian
if $d\mathcal{D}\left(\rho_{1}\left(t\right),\rho_{2}\left(t\right)\right)/dt$
becomes positive at any time $t$ and for some pair of initial states
$\rho_{1}\left(0\right)$ and $\rho_{2}\left(0\right)$~\citep{Breuer2009},
thus being a non-Markovian witness. Conversely, this means that, in
a Markovian dynamics, the distinguishability between any pair of states
always decreases monotonically in time, i.e., the derivative of the
trace distance is never positive. In other words, in a non-Markovian
dynamics there always exists some pair of initial states for which
the distinguishability (information) increases at a given time. In
particular, it is sufficient to assume a pair of initially pure and
orthogonal states to witness the non-Markovianity of a qubit system~\citep{Wi=0000DFmann2012}.
For that reason, we consider such a pair of states to be the eigenstates
of $\sigma_{z}$, i.e., the states $\Ket{0}$ and $\Ket{1}$, in order
to determine which parameter regimes induce a non-Markovian dynamics
into the refrigerant substance {[}see Fig.~\ref{Fig01}(b){]}.

We assume that the refrigerant and auxiliary qubits are resonant,
$2\pi\omega^{S}=2\pi\omega^{A}=2.2\text{ kHz}$, the vacuum decay
rate $\kappa=20\text{ Hz}$, and the temperature $T=2\hbar\omega^{A}/k_{\text{B}}\approx6\text{ pK}$
for the bosonic modes. These parameters are experimentally achievable
in NMR setups~\citep{Batalhao2014,Batalhao2015,Camati2016,Micadei2019}.
Using these parameters, in Fig.~\ref{Fig01}(b), we show for which
values of the ratio $J/\kappa$ the Markovian and non-Markovian regimes
are achieved. When $J/\kappa=40$ and $10$ the derivative
of the trace distance becomes positive and hence the system dynamics
is non-Markovian. On the other hand, for $J/\kappa=0.5$, we numerically
considered 10000 pairs of initial states where one of the orthogonal
states was varied throughout the north hemisphere of the Bloch sphere.
We found that the derivative of the trace distance never becomes positive,
hence we can suppose that the dynamics is Markovian. The plot in Fig.~\ref{Fig01}(b)
shows one representative example of these curves for the mentioned
pair of initial states. As shown in Ref.~\citep{Popovic2018}, which
studied a similar structured environment, the asymptotic state for
the system and the auxiliary qubit is a correlated one and the reduced
system state is a Gibbs state with an effective inverse temperature
$\beta_{c}^{\text{eff}}$ that is different from the inverse temperature
of the Markovian heat reservoir. 

\subsection{Quantum Otto refrigerator}

Let us consider a quantum Otto refrigerator with a single-qubit refrigerant
and the cycle of which is comprised by two driven adiabatic (no heat exchange)
and two undriven thermalization strokes. In an Otto refrigerator with
Markovian reservoirs, work and heat exchanges are associated with
the adiabatic and undriven thermalization strokes, respectively. These
thermodynamic quantities can be obtained through the difference between
the final and initial internal energies for a given stroke. The internal
energy at time $t$ is given by $U_{t}=\text{Tr}\left[\rho_{t}H\left(t\right)\right]$,
where $\rho_{t}$ is the density operator of the refrigerant (system)
and $H\left(t\right)$ is the Hamiltonian at a time $t$.

The refrigerant starts the cycle in the hot Gibbs state $\rho_{0}=\rho_{0}^{\text{eq,}h}=e^{-\beta_{h}H_{0}}/Z_{0}^{h}$,
where $\beta_{h}=1/k_{\text{B}}T_{h}$ is the hot inverse temperature,
$H_{0}$ is the initial Hamiltonian, and $Z_{0}^{h}=\text{Tr}\left[e^{-\beta_{h}H_{0}}\right]$
is the associated partition function. 

In the first stroke a compression is performed on the refrigerant
frequency, with driven Hamiltonian given by

\begin{equation}
H^{\text{com}}\left(t\right)=\frac{\hbar\omega\left(t\right)}{2}\sigma_{z},\label{driving}
\end{equation}
where the frequency is changed as a linear ramp as $\omega\left(t\right)=\left(1-t/\tau_{1}\right)\omega_{0}+\left(t/\tau_{1}\right)\omega_{\tau_{1}}$,
with $\omega_{0}$ and $\omega_{\tau_{1}}$ ($\omega_{0}>\omega_{\tau_{1}}$)
being the initial and final frequency of the refrigerant, respectively.
Hence the initial Hamiltonian is $H_{0}=H^{\text{com}}\left(0\right)=\left(\hbar\omega_{0}/2\right)\sigma_{z}$,
where ``com'' stands for compression. Although this stroke is performed
for a finite time, the reduced density operator of the system will
not present coherence in the energy eigenbasis. This is especially
due to the structure of the driving in Eq.~(\ref{driving}) which
commutes at different times, i.e., $\left[H^{\text{com}}\left(t\right),H^{\text{com}}\left(t'\right)\right]=0$
for $t\neq t'$. For recent studies of non-commutative driving in
quantum Otto heat engines, we refer to Refs.~\citep{Camati2019,kosloff2019A}.
In particular, in Ref.~\citep{Camati2019}, a quantum Otto heat engine
that generates coherence in the energy basis due to the non-commutativity
of the Hamiltonian was considered. It was shown that the effect of
coherence in a finite-time Otto cycle induces fast oscillations in
the figures of merit for the engine performance. In order to focus
on the non-Markovian effects due to the engineered cold reservoir,
we considered the commutative driving Hamiltonian in Eq.~(\ref{driving}).
The state at the end of the first stroke is given by $\rho_{\tau_{1}}=\mathcal{U}_{\tau_{1},0}\rho_{0}^{\text{eq,}h}\mathcal{U}_{\tau_{1},0}^{\dagger}$,
where $\mathcal{U}_{t,0}=\mathcal{T}_{>}\text{exp}\left[-\left(i/\hbar\right)\int_{0}^{t}ds\,H^{\text{\text{com}}}\left(s\right)\right]$
is the time-evolution operator, $\mathcal{T}_{>}$ is the time-ordering
operator, and $t\in\left[0,\tau_{1}\right]$. The final Hamiltonian
is $H^{\text{com}}\left(\tau_{1}\right)=\left(\hbar\omega_{\tau_{1}}/2\right)\sigma_{z}$,
such that the work performed on the refrigerant is given by $\left\langle W_{1}\right\rangle =U_{\tau_{1}}-U_{0}$.

In the second stroke the refrigerant system interacts with the engineered
cold reservoir described in the previous section, the Markovian part of which
has an inverse temperature $\beta_{c}=1/k_{\text{B}}T_{c}$. During
this stroke, the refrigerant Hamiltonian is kept fixed at $H^{c}\left(t\right)=\left(\hbar\omega_{\tau_{1}}/2\right)\sigma_{z}$
along the time $t\in\left[\tau_{1},\tau_{2}\right]$. For further
reference, we denote by $\Delta\tau_{c}=\tau_{2}-\tau_{1}$ the duration
of the second stroke. Moreover, some recent developments have pointed
out that incomplete (or partial) thermalization can be used to reach
a better performance in quantum heat engines~\citep{Camati2019,Renner2019}.
For that reason we consider an incomplete thermalization, denoting
by $\rho_{\tau_{2}}$ the state at the end of this stroke. Furthermore,
the structure of the engineered cold reservoir will be set to generate
a Markovian or non-Markovian dynamics on the refrigerant substance.
This choice is adjusted depending on the ratio of the coupling constant
$J$ between the refrigerant and the auxiliary qubit and the internal
coupling $\kappa$ between the auxiliary qubit and the cold Markovian
reservoir as considered in Fig.~\ref{Fig01}(b). Since $\kappa$
should be small (weak coupling) the regimes are essentially obtained
by varying $J$. The heat absorbed by the quantum refrigerant is given
by $\left\langle Q_{c}\right\rangle =U_{\tau_{2}}-U_{\tau_{1}}$.

In the third stroke, the refrigerant is decoupled from the structured
cold reservoir and its frequency is increased from $\omega_{\tau_{1}}$
to $\omega_{0}$ by means of an adiabatic expansion. The driving Hamiltonian
for this stroke is $H^{\text{exp}}\left(t\right)=H^{\text{com}}\left(\tau_{1}+\tau_{2}-t\right)$
applied for the time interval $t\in\left[\tau_{2},\tau_{3}\right]$,
where ``exp'' stands for expansion. The final state is $\rho_{\tau_{3}}=\mathcal{V}_{\tau_{3},\tau_{2}}\rho_{2}\mathcal{V}_{\tau_{3},\tau_{2}}^{\dagger}$
where $\mathcal{V}_{\tau_{3},\tau_{2}}=\mathcal{T}_{>}\text{exp}\left[-\left(i/\hbar\right)\int_{\tau_{2}}^{\tau_{3}}dt\:H^{\text{exp}}\left(t\right)\right]$
is the evolution operator associated to the adiabatic expansion. The
time duration of the third stroke is assumed to be the same as for
the first stroke, i.e., $\tau_{3}-\tau_{2}=\tau_{1}$, so that $H^{\text{exp}}\left(\tau_{3}\right)=H^{\text{com}}\left(0\right)$.
The work performed on the refrigerant during this stroke is $\left\langle W_{3}\right\rangle =U_{\tau_{3}}-U_{\tau_{2}}$.
At this point we can define the total work performed on the refrigerant
as $\left\langle W_{\text{net}}\right\rangle =\left\langle W_{1}\right\rangle +\left\langle W_{3}\right\rangle $.

Finally, in the fourth stroke and in order to close the cycle, this
last stroke comprises a complete thermalization of the refrigerant
with the hot reservoir at inverse temperature $\beta_{h}$. During
this process the Hamiltonian is kept fixed at $H^{\text{exp}}\left(\tau_{3}\right)=\left(\hbar\omega_{0}/2\right)\sigma_{z}$
along the time $t\in\left[\tau_{3},\tau_{4}\right]$. Once the condition
$\tau_{4}-\tau_{3}\gg\tau_{\text{rel}}^{h}$ is fulfilled, where $\tau_{\text{rel}}^{h}$
is the relaxation time for the hot Markovian heat reservoir, the final
state reaches $\rho_{\tau_{4}}=\rho_{0}^{\text{eq,}h}$. The heat
absorbed by the refrigerant is given by $\left\langle Q_{h}\right\rangle =U_{\tau_{4}}-U_{\tau_{3}}$.
The total time duration of a single cycle will be given by $\tau_{\text{cycle}}=\tau_{4}=\tau_{1}+\Delta\tau_{c}+\tau_{1}+\Delta\tau_{h}$.

Before we move forward to the discussion of the performance, we define
some other quantities that will turn out to be relevant. In the second
stroke, the first law of thermodynamics implies that $\left\langle Q_{c}^{S}\right\rangle +\left\langle Q_{c}^{R}\right\rangle +\Delta V_{c}^{SR}=0$,
where $-\left\langle Q_{c}^{R}\right\rangle $ is the energy released
by the engineered cold reservoir $R$, and $\Delta V_{c}^{SR}$ is
the change in the interaction energy between $S$ and $R$. The total
energy released by the engineered cold reservoir is, therefore, given
by the expression 
\begin{equation}
-\left\langle Q_{c}^{R}\right\rangle =\left\langle Q_{c}^{S}\right\rangle +\Delta V_{c}^{SR}.\label{eq: energy engineered cold reservoir}
\end{equation}
For our model, the auxiliary qubit couples weakly to the Markovian
heat reservoir, hence we have $\Delta V_{c}^{SR}=\Delta V_{c}^{SA}$,
where $\Delta V_{c}^{SA}=\text{Tr}\left[\rho_{\tau_{2}}^{SA}H_{\text{int}}^{SA}\right]-\text{Tr}\left[\rho_{\tau_{1}}^{SA}H_{\text{int}}^{SA}\right]$
is the interaction energy of the $SA$ composite system during the
second stroke. 

\section{Performance of the quantum Otto refrigerator\label{sec:Injected-power-and}}

The purpose of a refrigerator is to cool down the cold reservoir,
hence the quantity that has to fundamentally be considered to assess
the performance of a refrigerator is $-\left\langle Q_{c}^{R}\right\rangle $,
i.e., the energy released by the cold reservoir. Therefore, the COP
is defined by 
\begin{equation}
\epsilon=\frac{-\left\langle Q_{c}^{R}\right\rangle }{\left\langle W_{\text{net}}^{S}\right\rangle }=\frac{\left\langle Q_{c}^{S}\right\rangle +\Delta V_{c}^{SR}}{\left\langle W_{\text{net}}^{S}\right\rangle },
\end{equation}
where we used Eq.~(\ref{eq: energy engineered cold reservoir}) to
write the last equality. Similarly, the cooling power will be given
by $\left\langle \Gamma\right\rangle =-\left\langle Q_{c}^{R}\right\rangle /\tau_{\text{cycle}}$
while the injected power is $\left\langle \mathcal{P}\right\rangle =\left\langle W_{\text{net}}^{S}\right\rangle /\tau_{\text{cycle}}$.
When the interaction energy is small compared to the heat absorbed
by the system, i.e., $\left\langle Q_{c}^{S}\right\rangle +\Delta V_{c}^{SR}\approx\left\langle Q_{c}^{S}\right\rangle $,
we can see that the usual expressions for the figures of merit of a
refrigerator are recovered. One such a situation happens in classical
thermodynamics, where the coupling between the refrigerant and the
reservoir is small (weak-coupling regime). Not only that, the refrigerant
is a macroscopic system (a fluid or a gas) so that the internal energy, which scales with the volume, is larger than the interaction energy, which 
scales with the area (boundary). On the other hand, if the refrigerant
is strongly coupled to the reservoir, the interaction energy starts
playing a role in the thermodynamics~\citep{Newman2017,Newman2019}
and in the performance analysis of the refrigerator.

We obtain the following analytical expressions for any quantum refrigerant
(and not only for a qubit), as described in~\ref{sec:COP}. We later
restrict to our specific model again to present our numerical analysis.
The COP defined in Eq.~(\ref{eq:COP}) can be written as 
\begin{equation}
\epsilon=\gamma\frac{\epsilon_{\text{Carnot}}}{1+\epsilon_{\text{Carnot}}\mathcal{L}},\label{eq:COP}
\end{equation}
where $\epsilon_{\text{Carnot}}=\left(\beta_{c}/\beta_{h}-1\right)^{-1}$
is the Carnot COP and 
\begin{equation}
\mathcal{L}=\frac{D\left(\rho_{\tau_{1}}||\rho_{\tau_{1}}^{\text{eq,}c}\right)-D\left(\rho_{\tau_{2}}||\rho_{\tau_{2}}^{\text{eq,}c}\right)+D\left(\rho_{\tau_{3}}||\rho_{\tau_{3}}^{\text{eq,}h}\right)}{\beta_{h}\langle Q_{c}\rangle}\label{eq:COP lag}
\end{equation}
is the COP lag, previously defined for quantum engines~\citep{Camati2019,Peterson2018,Campisi2016}.
The multiplicative term in Eq.~(\ref{eq:COP}) is given by 

\begin{equation}
\gamma=\left(1+\frac{\Delta V_{c}^{SR}}{\left\langle Q_{c}^{S}\right\rangle }\right),\label{eq:degradation factor}
\end{equation}
and we call it the interaction-energy parameter. Similarly, the cooling
power can be written as 
\begin{equation}
\left\langle \Gamma\right\rangle =\gamma\frac{\left\langle Q_{c}^{S}\right\rangle }{\tau_{\text{cycle}}}.
\end{equation}
From these two equations we can see that the parameter $\gamma$ quantifies
how the COP and the cooling power change in the presence of a non-negligible
interaction energy between the refrigerant and the engineered cold
reservoir. 

In order to remain in the operation regime of a refrigerator, the
thermodynamic quantities must satisfy the constraints $-\left\langle Q_{c}^{R}\right\rangle >0$,
$\left\langle W_{\text{net}}\right\rangle >0$, and $\left\langle Q_{h}^{R}\right\rangle <0$.
These constraints imply that the interaction-energy parameter must
be positive, i.e., $\gamma>0$ (see~\ref{sec:COP}). For our specific
refrigerator cycle, the COP can be also written as $\epsilon=\gamma\epsilon_{\text{Otto}}$,
where $\epsilon^{\text{Otto}}=\omega_{\tau_{1}}/\left(\omega_{0}-\omega_{\tau_{1}}\right)$
(see~\ref{sec:COP Otto}). Therefore, the interaction-energy parameter
should be upper bounded by $1$ so that the COP is smaller than the
Otto COP. For a wide range of parameters, we verified numerically
that our refrigerator does not surpass the Otto COP. If one assess
the performance of the refrigerator ignoring this parameter $\gamma$,
either by simply not accounting for it or because one might not be
aware if the interaction energy is not negligible, the performance
of the refrigerator will be overestimated by precisely $1/\gamma$.
More specifically, ignoring a parameter of $\gamma=0.9$, the COP
and the cooling power would be overestimated by about $11\%$, meaning
that the computed COP and cooling power will be $11\%$ larger than
the true COP and cooling power.

Before we present our numerical analysis, we note that we have chosen
a driving Hamiltonian that commutes at different times, $\left[H^{\text{com}}\left(t\right),H^{\text{com}}\left(t'\right)\right]=0$.
This implies an Otto cycle without quantum friction~\citep{Feldmann2006,Camati2019,Batalhao2015,Feldmann01,Feldman02,Campo01,Thomas01,Plastina01,Paternostro2014}.
We assume this dynamics for the driving Hamiltonian in order to solely
focus on the non-Markovian aspects of the Otto cycle. Since our refrigerator
has no quantum friction, in the Markovian regime where the interaction
energy is negligible, $\gamma=1$, the refrigerator reaches the Otto
COP, $\epsilon=\epsilon_{\text{Otto}}$. In the case of the non-Markovian
regime, even in the absence of quantum friction, the Otto COP will
not be generally reached, due to the presence of the interaction energy. 

In the following, we performed a numerical simulation of our quantum
Otto refrigerator, assuming energy scales compatible with quantum
thermodynamics experiments performed in NMR setups~\citep{Peterson2018,Batalhao2014,Batalhao2015,Camati2016,Micadei2019}.
The initial and final gaps of the compression stroke are chosen as
$\omega_{0}/2\pi=3.6\text{ kHz}$ and $\omega_{\tau_{1}}/2\pi=2.2\text{ kHz}$,
respectively. The cold ($T_{c}$) and hot ($T_{h}$) temperatures
are chosen such that $T_{c}=1/\left(2.5\omega_{\tau_{1}}\right)$
and $T_{h}=1/\left(2.5\omega_{0}\right)$ with inverse temperatures
given by $\beta_{c}=1/T_{c}$ and $\beta_{h}=1/T_{h}$. Finally, we
assume that the vacuum decay rates of the cold and hot Markovian reservoirs
are $\gamma_{c}=\gamma_{h}=20\text{ Hz}$, where $\gamma_{c}=\kappa$
from Sec.~\ref{subsec:Model-of-structured}.

\begin{figure}[h]
\begin{centering}
\includegraphics[scale=0.65]{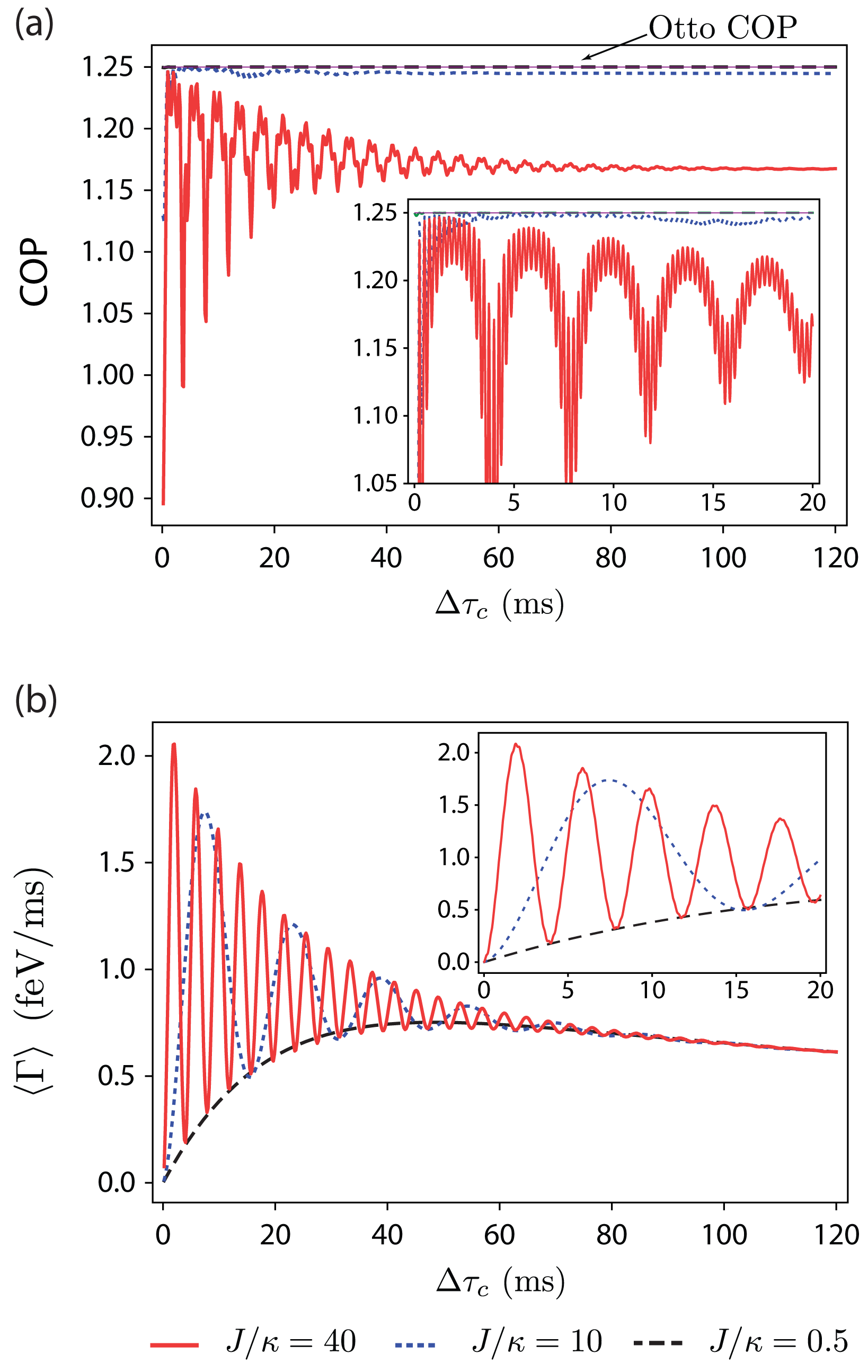}
\par\end{centering}
\caption{COP (a) and cooling power (b) as a function of the time of the second
stroke $\Delta\tau_{c}$ for three values of the ratio $J/\kappa$:
40 (solid red line), 10 (dotted blue line), and 0.5 (dashed black
line). The oscillations of the COP are very fast, as seen in the inset
of (a), and hence the oscillations seen in (a) represent just a coarse graining
of the actual oscillations. From both figures we can see that the
larger the ratio $J/\kappa$ is, which implies a larger effect of
the non-Markovian dynamics, the larger the effect on the performance
as well, namely, larger oscillation amplitudes and frequency of oscillations.
For the Markovian regime ($J/\kappa=0.5$), there is no oscillation
in the cooling power, and the COP is the Otto COP, as expected from
the fact that our cycle has no quantum friction. We considered $\tau_{1}=0.75\text{ ms}$.
\label{fig:COP and cooling rate during stroke 2}}
\end{figure}

We see in Fig.~\ref{fig:COP and cooling rate during stroke 2} that
the Markovian regime (dashed black line) reproduces the expected results.
The COP reaches the Otto limit, since there is no quantum friction
in our cycle. Furthermore, the cooling power $\langle\Gamma\rangle$
decreases for second strokes of long duration, for both the non-Markovian
and the Markovian regimes. We can clearly see the effect of the non-Markovian
dynamics on both the COP and the cooling power, which is to induce
oscillations of different amplitudes and frequencies, depending on
the ratio $J/\kappa$. For a larger ratio of $J/\kappa$ {[}solid
red line in Fig.~\ref{fig:COP and cooling rate during stroke 2}(b){]}
the cooling power oscillates quite more frequently but does not have
a maximum value that exceeds considerably the maximum of  a moderate
ratio of $J/\kappa$ {[}dotted blue line in Fig.~\ref{fig:COP and cooling rate during stroke 2}(b){]}.
In the non-Markovian regime, the oscillation frequency increases considerably
meaning that a better control of the allocation time on the second
stroke is needed to end up in the peak of the cooling power curves.
On the other hand, in Fig.~\ref{fig:COP and cooling rate during stroke 2}(a),
a larger ratio of $J/\kappa$ (solid red line) degrades considerably
the COP and the timescale for the oscillations are much smaller than
for the cooling power {[}compare the insets in Figs.~\ref{fig:COP and cooling rate during stroke 2}(a)
and \ref{fig:COP and cooling rate during stroke 2}(b){]}. This means
that, for larger values of $J/\kappa$ an extremely good amount of
control should be necessary to make the refrigerator operate in the
maximum of the corresponding COP. We also see that, for a moderate
ratio of $J/\kappa$ (dotted blue line), although the COP is smaller
than the Otto COP it remains very close to it throughout the time range
considered in the plot. In other words, the value of the COP for moderate
values of $J/\kappa$ does not change much from its asymptotic value
(reached for long time durations of the second stroke $\Delta\tau_{c}$).
Therefore, for our model of an engineered cold reservoir, we can conclude
that a moderate non-Markovian dynamics ($J/\kappa=10$) performs generally
better than a Markovian ($J/\kappa=0.5$) or a strong non-Markovian
($J/\kappa=40$) regime. Comparing to the Markovian regime, the moderate
non-Markovian regime has almost the same COP and improves considerably
on the cooling power, whereas, comparing the moderate (dotted blue
line) with the strong (solid red line) non-Markovian regime, the COP
is larger while the cooling power oscillates with a smaller frequency,
making it easier to adjust the cycle time to operate in the optimal
regimes.

\begin{figure}[h]
\begin{centering}
\includegraphics[scale=0.65]{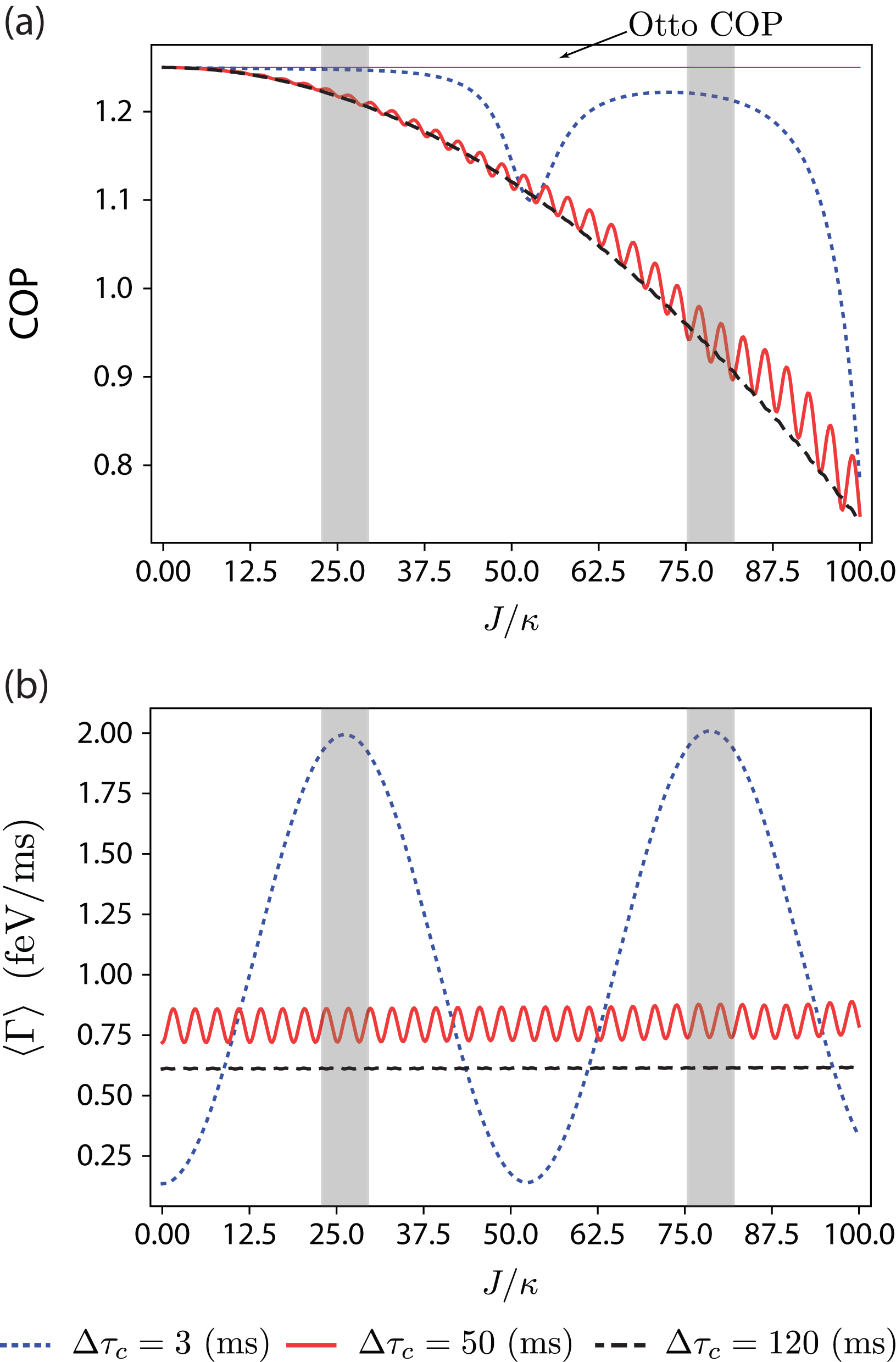}
\par\end{centering}
\caption{COP (a) and cooling power (b) as a function of the ratio $J/\kappa$
for three values of the duration of the second stroke $\Delta\tau_{c}$:
3 ms (dotted blue line), 50 ms (solid red line), and 120 ms (dashed
black line). The Markovian regime corresponds to values of about or
smaller than $J/\kappa=0.5$. The gray area on both figures shows
the region of $J/\kappa$ values for which the cooling power is close
to its maximum. We considered $\tau_{1}=0.75\text{ ms}$. \label{fig:COP and cooling rate as function of J}}
\end{figure}

We now fix three values for the duration of the second stroke $\Delta\tau_{c}$
and analyze how the performance of the refrigerator changes with the
ratio $J/\kappa$ (see Fig.~\ref{fig:COP and cooling rate as function of J}).
For a long second stroke, $\Delta\tau_{c}=120\text{ ms}$ (dashed
black line), the refrigerator and auxiliary qubits practically reach
their asymptotic states. We see that the COP degrades very quickly with
increasing $J/\kappa$, and that in the Markovian regime ($J/\kappa\apprle0.5$)
it reaches the Otto COP [see dashed black line in Fig.~\ref{fig:COP and cooling rate as function of J}(a)].
The cooling power, on the other hand, stays constant, showing that
in the asymptotic regime the non-Markovian regime is no better or
worse than the Markovian regime [see dashed black line in Fig.~\ref{fig:COP and cooling rate as function of J}(b)].
For a moderate duration of the second stroke, the COP still decreases
quickly but with an oscillating profile, whereas the cooling power
also does not represent a big difference between the non-Markovian
and Markovian regimes [see solid red lines in Figs.~\ref{fig:COP and cooling rate as function of J}(a)
and \ref{fig:COP and cooling rate as function of J}(b)]. As expected,
for a very short duration of the second stroke [see dotted blue lines
in Figs.~\ref{fig:COP and cooling rate as function of J}(a) and
\ref{fig:COP and cooling rate as function of J}(b)] and for the Markovian
regimes, the COP is the Otto limit and the cooling power is close
to zero, since the refrigerant has not much time to absorb the energy
from the cold reservoir. When the ratio $J/\kappa$ starts to get
larger, the non-Markovian dynamics starts to impact more strongly.
The cooling power starts increasing and then oscillates. It is
interesting to note that, even though the COP is strictly smaller than
the Otto COP for the non-Markovian regime, it stays very close to
the Otto COP for a reasonable portion of the initial values of $J/\kappa$.
Then, there is a brief decay in the COP which starts increasing again
and stays almost constant before decreasing abruptly. This behavior
is notably different than for a moderate duration of the second stroke.
We highlighted in gray the range of the $J/\kappa$ parameter for
which the cooling power is close to its maximum. For these shaded
regions (in Figs. \ref{fig:COP and cooling rate as function of J}(a)
and \ref{fig:COP and cooling rate as function of J}(b)), the COP
is about 99.8\% and about 97.2\% of the Otto limit, for the first
and second highlighted region from left to right, respectively. These
regions point out the most interesting parameter regimes in which
the refrigerator should be operated, among the three plotted curves.
Figures \ref{fig:COP and cooling rate as function of J}(a) and \ref{fig:COP and cooling rate as function of J}(b)
show also how the short duration of the second stroke makes a huge
difference in the performance of the non-Markovian refrigerator.

\begin{figure}[H]
\begin{centering}
\includegraphics[scale=0.65]{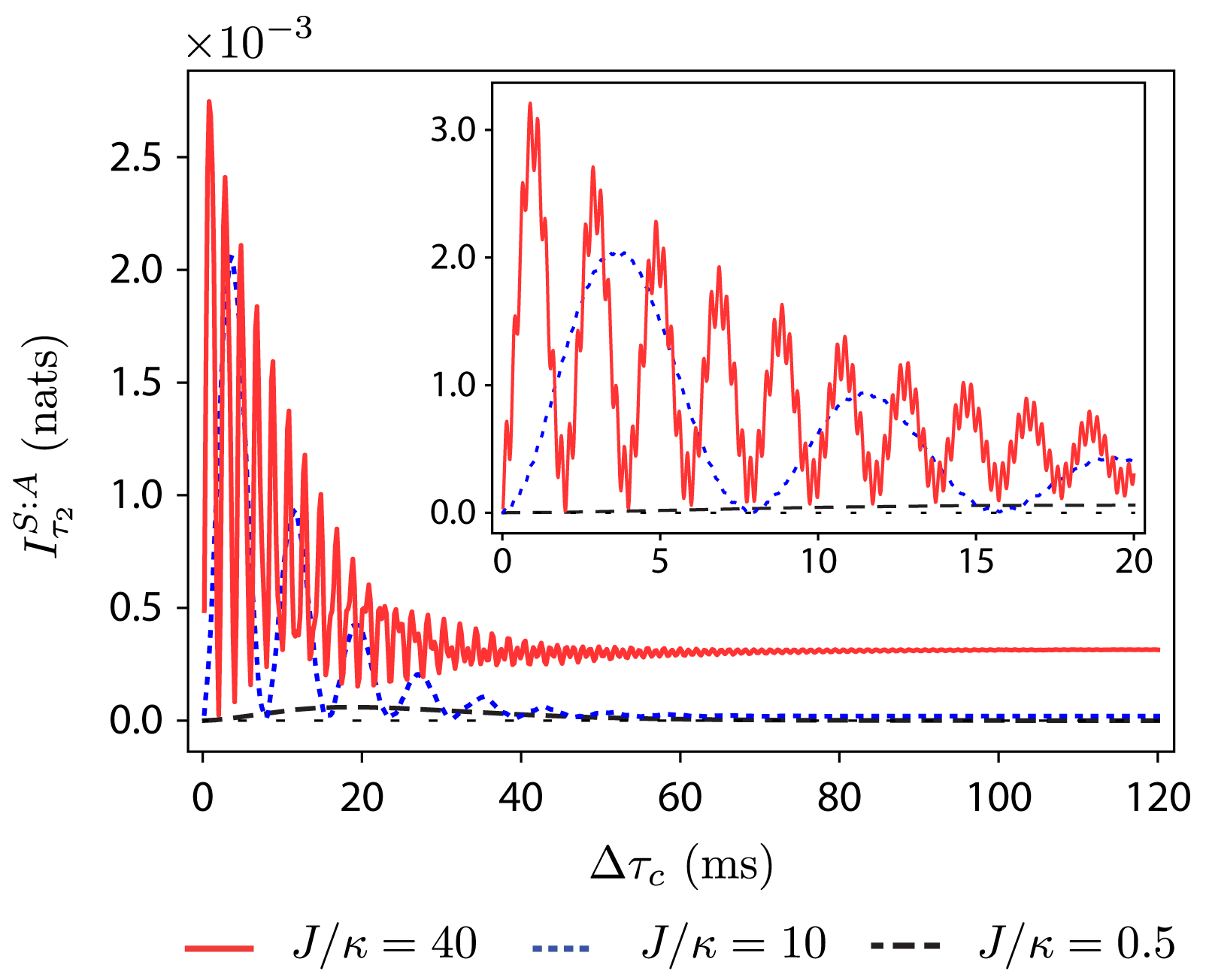}
\par\end{centering}
\caption{Mutual information $I_{\tau_{2}}^{S:A}$ between the refrigerant and
the auxiliary qubit at the end of the second stroke. For the strong
non-Markovian regime, the solid red line with $J/\kappa=40$, the mutual
information oscillates very rapidly (see the inset). The oscillatory
behavior of the plot is a coarse graining of the true oscillations.
We can see that in the asymptotic regime, for a long duration of the
second stroke, the $SA$ system reaches a correlated state with given
mutual information. In the Markovian regime the correlations are zero
(dashed black line) while for a moderate non-Markovian regime (dotted
blue line) it oscillates less frequently and almost reaches zero at
the end. We considered $\tau_{1}=0.75\text{ ms}$ and $1\text{ nat }=1/\ln2\approx1.44\text{ bit}$.
\label{fig:mutual information}}
\end{figure}

We finish our discussion in this section by showing how the mutual
information between the refrigerant and the auxiliary qubit changes
in the second stroke (see Fig.~\ref{fig:mutual information}). The
mutual information is given by $I_{\tau_{2}}^{S:A}=S\left(\rho_{\tau_{2}}^{S}\right)+S\left(\rho_{\tau_{2}}^{A}\right)-S\left(\rho_{\tau_{2}}^{SA}\right)$,
where $\rho_{\tau_{2}}^{SA}$, $\rho_{\tau_{2}}^{S}$, and $\rho_{\tau_{2}}^{A}$
are the composite $SA$, the refrigerant $S$, and the auxiliary qubit
$A$ states at the end of the second stroke. The quantity $S\left(\rho\right)=-\text{Tr}\left[\rho\ln\rho\right]$
is the von Neumann entropy. We see that, for $J/\kappa=40$ (solid
red line), the mutual information oscillates very rapidly, reaching
a finite value at the end, showing that the asymptotic $SA$ state
is correlated~\citep{Newman2017}. For a moderate non-Markovian dynamics,
$J/\kappa=10$ (dotted blue line), the mutual information also oscillates,
but with a smaller frequency, reaching almost zero at the end. For
the Markovian regime, $J/\kappa=0.5$ (dashed black line), the mutual
information is almost zero everywhere. Comparing the oscillations
in the inset of Fig.~\ref{fig:mutual information} and those of the
inset of Fig.~\ref{fig:COP and cooling rate during stroke 2}(b),
we see that the peaks of the cooling power are not associated directly
to the peaks of the mutual information. The same conclusion applies
comparing the mutual information with the inset of Fig.~\ref{fig:COP and cooling rate during stroke 2}(a).
This shows that, although the presence of mutual information is important
because it is related to the non-Markovian dynamics, the improvement
of the refrigerator performance is not directly related to the correlation
created between the refrigerant and the engineered cold reservoir.

\section{Conclusions\label{sec:Conclusions}}

We analyzed the performance of a quantum Otto refrigerator the refrigerant
of which interacts with an engineered cold reservoir, comprised by a Markovian
reservoir and an auxiliary two-level system. Depending on the coupling
between the refrigerant and the structured cold environment the dynamics
of the refrigerant can be either Markovian or non-Markovian, changing
the performance of the refrigerator. In our model, the non-Markovian
regime is reached when the interaction between the refrigerant and
the reservoir is strong, in the sense that the contribution of the
interaction energy between the refrigerant and the engineered reservoir
cannot be neglected. Concisely, the heat removed from the cold reservoir
(the prime purpose of a refrigerator) is not the heat absorbed by
the refrigerant, because the interaction Hamiltonian stores a nontrivial
amount of energy.

Taking the interaction energy into account, we defined the figures
of merit for the refrigerator: the COP, the cooling power, and the
injected power. We showed that the interaction-energy contribution
can be recast as a parameter $\gamma$, given by Eq.~(\ref{eq:degradation factor}),
multiplying the usual expressions for the COP and cooling power, which
disregard such a contribution. These analytical expressions are valid
irrespective of the nature of the quantum refrigerant. In light of
these expressions, we can conclude that an overestimation of the refrigerator
performance would be made if such an interaction-energy contribution
was ignored in the analysis. In fact, it is in principle possible
to find an operation regime in which one might think the refrigerator
is working properly, because the refrigerant is absorbing energy,
but in fact it is not operating as a refrigerator at all, i.e.,
there is no energy being removed from the cold reservoir. The increase in the
refrigerant energy comes solely from the interaction energy.

Performing a numerical analysis, we observed the expected behavior
for the Markovian regime of operation, reached for a sufficiently
small coupling between the refrigerant and the engineered cold reservoir.
On the other hand, for the non-Markovian regimes of operation, we
see that both the COP and the cooling power start oscillating, a feature
coming from the backflow of information from the non-Markovian environment.
Both COP and cooling power oscillate faster as the coupling and the non-Markovian
effect increase, but the COP decreases significantly as well, while
the cooling power increases slightly for a small time of interaction
between them (duration of the second stroke). We conclude that, at
least for the model considered in our paper, a moderate non-Markovian
effect improves the performance of the refrigerator, if compared to
the Markovian or strong non-Markovian regimes. We also point out that
in order to exploit the non-Markovian effects it is important to operate
the refrigerator in a sufficiently short interaction timescale between
the refrigerant and the engineered cold reservoir.

The finite-time performance of the present quantum Otto refrigerator
may be enhanced by the information backflow provided one has sufficient
control over the time allocated in each stroke. The parameters considered
in the numerical simulation can be experimentally realized with current
technologies, for instance, in nuclear magnetic resonance setups.
Along with other studies addressing non-Markovian effects in quantum
thermodynamics, we hope that our analyses help to unveil the role
of memory effects in quantum thermal machines.

\section*{Acknowledgments}

We thank Ivan Medina and Wallace S. Teixeira for fruitful discussions
during the early stages of this work. We acknowledge financial support
from Universidade Federal do ABC (UFABC), Conselho Nacional de Desenvolvimento Científico e Tecnológico (CNPq), Coordenação de
Aperfeiçoamento de Pessoal de Nível Superior (CAPES), and Fundação de Amparo à Pesquisa do Estado de São
Paulo (FAPESP). This research was performed as
part of the Brazilian National Institute of Science and Technology
for Quantum Information (INCT-IQ). P. A. C. acknowledges CAPES and
Templeton World Charity Foundation, Inc (TWCF). This publication was made
possible through the support of the Grant No. TWCF0338 from TWCF. J.F.G.S acknowledges support
from FAPESP (Grant No. 19/04184-5).

\setcounter{section}{0}
\global\long\def\thesection{Appendix \Alph{section}}

\section{Master Equation of the Non-Markovian Model \label{sec:Master-Equation-of}}

\setcounter{figure}{0}
\setcounter{equation}{0}
\global\long\def\thefigure{A\arabic{figure}}
\global\long\def\theequation{A\arabic{equation}}

The master equation of the non-Markovian model employed is given by
Eq.~(\ref{eq:master equation}). Here, we provide some details on
the derivation of this master equation. We will follow the approach
of Refs.~\citep{Rivas2010,Rivas2012book}.

We begin considering the diagonalization of the two-qubit Hamiltonian
{[}Eq.~(\ref{eq:two qubit hamiltonian}){]}
\begin{equation}
H^{SA}=\sum_{i=S,A}\frac{\hbar\omega^{i}}{2}\sigma_{z}^{i}+\hbar J\sigma_{x}^{S}\sigma_{x}^{A}.
\end{equation}
The eigenvectors and eigenvalues are:
\begin{equation}
\Ket{E_{3}}=\alpha\Ket{00}+\xi\Ket{11}\text{ and }E_{3}=\frac{\hbar}{2}\sqrt{4J^{2}+\Omega^{2}};
\end{equation}
\begin{equation}
\Ket{E_{2}}=\eta\Ket{01}-\delta\Ket{10}\text{ and }E_{2}=\frac{\hbar}{2}\sqrt{4J^{2}+\Delta^{2}};
\end{equation}
\begin{equation}
\Ket{E_{1}}=\delta\Ket{01}+\eta\Ket{10}\text{ and }E_{1}=-E_{2};
\end{equation}
\begin{equation}
\Ket{E_{0}}=-\xi\Ket{00}+\alpha\Ket{11}\text{ and }E_{0}=-E_{3}.
\end{equation}
We introduced the parameters $\Delta=\omega^{S}-\omega^{A}$, $\Omega=\omega^{S}+\omega^{A}$,
\begin{equation}
\alpha=\frac{\Omega+\sqrt{4J^{2}+\Omega^{2}}}{\sqrt{4J^{2}+\left(\Omega+\sqrt{4J^{2}+\Omega^{2}}\right)^{2}}},
\end{equation}
\begin{equation}
\xi=\frac{2J}{\sqrt{4J^{2}+\left(\Omega+\sqrt{4J^{2}+\Omega^{2}}\right)^{2}}},
\end{equation}
\begin{equation}
\eta=\frac{\Delta+\sqrt{4J^{2}+\Delta^{2}}}{\sqrt{4J^{2}+\left(\Delta+\sqrt{4J^{2}+\Delta^{2}}\right)^{2}}},
\end{equation}
and
\begin{equation}
\delta=\frac{-2J}{\sqrt{4J^{2}+\left(\Delta+\sqrt{4J^{2}+\Delta^{2}}\right)^{2}}}.
\end{equation}

\begin{figure}
\begin{centering}
\includegraphics{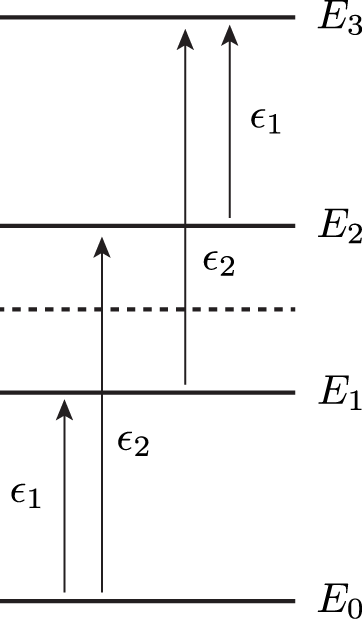}
\par\end{centering}
\caption{Transition frequencies. Among the six positive frequencies (from a
low- to a high-energy level) only four appear on the master equation
due to the structure of the interaction Hamiltonian. These are shown
in the energy diagram. \label{fig:Transition-frequencies}}

\end{figure}

The master equation of the two-qubit system is given by~\citep{Rivas2012book,Breuer2002book,Alicki2007book}
\begin{equation}
\frac{d}{dt}\rho_{t}^{SA}=-\frac{i}{\hbar}\left[H^{SA},\rho_{t}^{SA}\right]+\mathcal{D}\left(\rho_{t}^{SA}\right),
\end{equation}
where 
\begin{align}
\mathcal{D}\left(\rho_{t}^{SA}\right)= & \sum_{\omega\in\mathcal{F}}\sum_{k,l=1,2}\gamma_{kl}\left(\omega\right)\nonumber \\
 & \times\left[A_{l}\left(\omega\right)\rho_{t}^{SA}A_{k}^{\dagger}\left(\omega\right)-\frac{1}{2}\left\{ A_{k}^{\dagger}\left(\omega\right)A_{l}\left(\omega\right),\rho_{t}^{SA}\right\} \right]\label{eq:dissipator appendix}
\end{align}
is the dissipator superoperator. We note that we have disregarded
the Lamb-shift Hamiltonian since it contributes to an overall energy
shift. The set $\mathcal{F}$ is comprised by all, positive and negative,
transition frequencies $\omega_{nm}=\left(E_{m}-E_{n}\right)/\hbar$
for $m,n\in\left\{ 0,1,2,3\right\} $. Since the composite system
$SA$ is a four-level system, there are 12 transition frequencies,
6 positive and 6 negative, $\omega_{nm}$ and $\omega_{mn}=-\omega_{nm}$
with $n<m$, respectively. As we will explain below, among the 6 positive
(or negative) transition frequencies only 2 doubly-degenerate ones
are relevant to our problem (see Fig.~\ref{fig:Transition-frequencies}).
The positive transition frequencies are $\omega_{01}=\omega_{23}=\epsilon_{1}$
and $\omega_{02}=\omega_{13}=\epsilon_{2}$, where 
\begin{equation}
\epsilon_{1}=\frac{1}{2}\left[\sqrt{\Omega^{2}+4J^{2}}-\sqrt{\Delta^{2}+4J^{2}}\right]
\end{equation}
and 
\begin{equation}
\epsilon_{2}=\frac{1}{2}\left[\sqrt{\Omega^{2}+4J^{2}}+\sqrt{\Delta^{2}+4J^{2}}\right].
\end{equation}

Before we proceed, we rewrite the interaction Hamiltonian {[}Eq.~(\ref{eq:total hamiltonian}){]}
of the two-qubits system and the heat reservoir as 
\begin{align}
V^{AR}= & \int_{0}^{\omega_{\text{max}}}d\omega\,h\left(\omega\right)\left(\sigma_{+}^{A}\otimes b_{\omega}+\sigma_{-}^{A}\otimes b_{\omega}^{\dagger}\right)\nonumber \\
= & A_{1}\otimes B_{1}+A_{2}\otimes B_{2},
\end{align}
where $\sigma_{\pm}=\left(\sigma_{x}\pm i\sigma_{y}\right)/2$ and
the continuum limit has been taken at the Hamiltonian level~\citep{Rivas2012book}.
The defined operators are: $A_{1}=\sigma_{x}^{A}$, $A_{2}=\sigma_{y}^{A}$,
$B_{1}=\int_{0}^{\omega_{\text{max}}}d\omega\,h\left(\omega\right)\frac{b_{\omega}^{\dagger}+b_{\omega}}{2}$,
and $B_{2}=\int_{0}^{\omega_{\text{max}}}d\omega\,h\left(\omega\right)\frac{ib_{\omega}-ib_{\omega}^{\dagger}}{2}$.
Using the relation $B_{i}=\int_{-\omega_{\text{max}}}^{\omega_{\text{max}}}d\omega\,B_{i}\left(\omega\right)$,
one can obtain the operators $B_{1}\left(\omega\right)=\frac{h\left(\omega\right)b_{\omega}}{2}$,\textbf{
$B_{1}\left(-\omega\right)=\frac{h\left(\omega\right)b_{\omega}^{\dagger}}{2}$},
$B_{2}\left(\omega\right)=i\frac{h\left(\omega\right)b_{\omega}}{2}$,
and $B_{2}\left(-\omega\right)=-i\frac{h\left(\omega\right)b_{\omega}^{\dagger}}{2}$.
Since there are two reservoirs operators, $k,l\in\left\{ 1,2\right\} $
in Eq.~(\ref{eq:dissipator appendix}) and, therefore, there are
four decay rates for each transition frequency $\omega\in\mathcal{F}$.

First, we find the decay rates for a given $\omega$, which are given
by the formula~\citep{Rivas2012book}
\begin{align}
\gamma_{kl}\left(\omega\right)= & \int_{-\infty}^{+\infty}ds\,e^{i\omega s}\text{Tr}_{R}\left[B_{k}\left(s\right)B_{l}\rho^{R,\text{eq}}\right]\nonumber \\
= & 2\pi\text{Tr}_{R}\left[B_{k}\left(\omega\right)B_{l}\rho^{R,\text{eq}}\right],
\end{align}
where $\omega\in\mathcal{F}$ and $\rho^{R,\text{eq}}$ is the Gibbs
state of the reservoir. Knowing that $\text{Tr}_{R}\left[a_{\omega}a_{\omega'}^{\dagger}\rho^{R,\text{eq}}\right]=\left[1+n_{\text{BE}}\left(\omega\right)\right]\delta\left(\omega-\omega'\right)$,
$\text{Tr}_{R}\left[a_{\omega}^{\dagger}a_{\omega'}\rho^{R,\text{eq}}\right]=n_{\text{BE}}\left(\omega\right)\delta\left(\omega-\omega'\right)$,
and $\text{Tr}_{R}\left[a_{\omega}a_{\omega'}\rho^{R,\text{eq}}\right]=\text{Tr}_{R}\left[a_{\omega}^{\dagger}a_{\omega'}^{\dagger}\rho^{R,\text{eq}}\right]=0$,
one can show for the positive transition frequencies $\omega\in\mathcal{F}_{+}=\left\{ \omega\in\mathcal{F}\,|\,\omega\geq0\right\} $
that
\begin{equation}
\gamma_{11}\left(\omega\right)=\frac{\pi}{2}J\left(\omega\right)\left[1+n_{\text{BE}}\left(\omega\right)\right],
\end{equation}
\begin{equation}
\gamma_{12}\left(\omega\right)=-i\frac{\pi}{2}J\left(\omega\right)\left[1+n_{\text{BE}}\left(\omega\right)\right],
\end{equation}
\begin{equation}
\gamma_{21}\left(\omega\right)=i\frac{\pi}{2}J\left(\omega\right)\left[1+n_{\text{BE}}\left(\omega\right)\right],
\end{equation}
and
\begin{equation}
\gamma_{22}\left(\omega\right)=\frac{\pi}{2}J\left(\omega\right)\left[1+n_{\text{BE}}\left(\omega\right)\right],
\end{equation}
where $n_{\text{BE}}\left(\omega\right)=\left(e^{\beta\hbar\omega}-1\right)^{-1}$
is the Bose-Einstein distribution and $J\left(\omega\right)=\left[h\left(\omega\right)\right]^{2}$
is the spectral density. On the other hand, for the negative transition
frequencies $\omega\in\mathcal{F}_{-}=\left\{ \omega\in\mathcal{F}\,|\,\omega<0\right\} $,
one finds
\begin{equation}
\gamma_{11}\left(-\omega\right)=\frac{\pi}{2}J\left(\omega\right)n_{\text{BE}}\left(\omega\right),
\end{equation}
\begin{equation}
\gamma_{12}\left(-\omega\right)=i\frac{\pi}{2}J\left(\omega\right)n_{\text{BE}}\left(\omega\right),
\end{equation}
\begin{equation}
\gamma_{21}\left(-\omega\right)=-i\frac{\pi}{2}J\left(\omega\right)n_{\text{BE}}\left(\omega\right),
\end{equation}
and
\begin{equation}
\gamma_{22}\left(-\omega\right)=\frac{\pi}{2}J\left(\omega\right)n_{\text{BE}}\left(\omega\right).
\end{equation}

Now, we need the operators $A_{k}\left(\omega\right)$, which come
from the system operators $\sigma_{x}^{A}$ and $\sigma_{y}^{A}$.
They are given by~\citep{Rivas2012book,Breuer2002book}
\begin{equation}
A_{k}\left(\omega_{nm}\right)=\sum_{E_{m}-E_{n}=\omega_{nm}}\Pi_{n}A_{k}\Pi_{m},
\end{equation}
where the projection operators are $\Pi_{n}=\Ket{E_{n}}\Bra{E_{n}}$.
Since $A_{1}=\mathds{1}^{S}\otimes\sigma_{x}^{A}$ and $A_{2}=\mathds{1}^{S}\otimes\sigma_{y}^{A}$
one can show that
\begin{equation}
\Pi_{n}A_{1}\Pi_{m}=\begin{pmatrix}0 & \alpha\eta-\delta\xi & -\left(\alpha\delta+\eta\xi\right) & 0\\
\alpha\eta-\delta\xi & 0 & 0 & \alpha\delta+\eta\xi\\
-\left(\alpha\delta+\eta\xi\right) & 0 & 0 & \alpha\eta-\delta\xi\\
0 & \alpha\delta+\eta\xi & \alpha\eta-\delta\xi & 0
\end{pmatrix}
\end{equation}
and \begin{widetext}

\begin{equation}
\Pi_{n}A_{2}\Pi_{m}=\begin{pmatrix}0 & i\left(\alpha\eta+\delta\xi\right) & -i\left(\alpha\delta-\eta\xi\right) & 0\\
-i\left(\alpha\eta+\delta\xi\right) & 0 & 0 & i\left(\alpha\delta-\eta\xi\right)\\
i\left(\alpha\delta-\eta\xi\right) & 0 & 0 & i\left(\alpha\eta+\delta\xi\right)\\
0 & -i\left(\alpha\delta-\eta\xi\right) & -i\left(\alpha\eta+\delta\xi\right) & 0
\end{pmatrix}.
\end{equation}
\end{widetext}Note how the operators $A_{k}\left(\omega\right)$
associated with the transition frequencies $\omega_{03}$, $\omega_{30}$,
$\omega_{12}$, $\omega_{21}$ are identically zero. That is why these
transition frequencies are not relevant, i.e., because their associated
operators are zero due to the structure of the interaction Hamiltonian.
Explicitly, one has 
\begin{equation}
A_{k}\left(\epsilon_{1}\right)=\Pi_{0}A_{k}\Pi_{1}+\Pi_{2}A_{k}\Pi_{3},
\end{equation}
\begin{equation}
A_{k}\left(\epsilon_{2}\right)=\Pi_{0}A_{k}\Pi_{2}+\Pi_{1}A_{k}\Pi_{3},
\end{equation}
$A_{k}\left(-\epsilon_{1}\right)=A_{k}^{\dagger}\left(\epsilon_{1}\right)$,
and $A_{k}\left(-\epsilon_{2}\right)=A_{k}^{\dagger}\left(\epsilon_{2}\right)$,
for $k\in\left\{ 1,2\right\} $. In summary, there are four system
operators, two for each nondegenerate transition frequency $\epsilon_{1}$
and $\epsilon_{2}$. This means that there will be four dissipative
channels in the master equation {[}see Eq.~(\ref{eq:master equation}){]}.
Replacing the decay rates $\gamma_{kl}\left(\omega\right)$ and system
operators $A_{k}\left(\omega\right)$, one obtains the master equation
employed, whose dissipator is given by {\small{}
\begin{align}
{\normalcolor \mathcal{D}\left(\rho_{t}^{SA}\right)=} & {\normalcolor \gamma^{\downarrow}\left(\epsilon_{1}\right)\left[L_{1}\left(\epsilon_{1}\right)\rho_{t}^{SA}L_{1}^{\dagger}\left(\epsilon_{1}\right)-\frac{1}{2}\left\{ L_{1}^{\dagger}\left(\epsilon_{1}\right)L_{1}\left(\epsilon_{1}\right),\rho_{t}^{SA}\right\} \right]}\nonumber \\
{\normalcolor } & {\normalcolor +\gamma^{\downarrow}\left(\epsilon_{2}\right)\left[L_{2}\left(\epsilon_{2}\right)\rho_{t}^{SA}L_{2}^{\dagger}\left(\epsilon_{2}\right)-\frac{1}{2}\left\{ L_{2}^{\dagger}\left(\epsilon_{2}\right)L_{2}\left(\epsilon_{2}\right),\rho_{t}^{SA}\right\} \right]}\nonumber \\
{\normalcolor } & {\normalcolor +\gamma^{\uparrow}\left(\epsilon_{1}\right)\left[L_{1}^{\dagger}\left(\epsilon_{1}\right)\rho_{t}^{SA}L_{1}\left(\epsilon_{1}\right)-\frac{1}{2}\left\{ L_{1}\left(\epsilon_{1}\right)L_{1}^{\dagger}\left(\epsilon_{1}\right),\rho_{t}^{SA}\right\} \right]}\nonumber \\
{\normalcolor } & {\normalcolor +\gamma^{\uparrow}\left(\epsilon_{2}\right)\left[L_{2}^{\dagger}\left(\epsilon_{2}\right)\rho_{t}^{SA}L_{2}\left(\epsilon_{2}\right)-\frac{1}{2}\left\{ L_{2}\left(\epsilon_{2}\right)L_{2}^{\dagger}\left(\epsilon_{2}\right),\rho_{t}^{SA}\right\} \right],}
\end{align}
}where $\gamma^{\uparrow}\left(\epsilon_{k}\right)=\frac{\pi}{2}J\left(\epsilon_{k}\right)n_{\text{BE}}\left(\epsilon_{k}\right)$,
$\gamma^{\downarrow}\left(\epsilon_{k}\right)=\frac{\pi}{2}J\left(\epsilon_{k}\right)\left[1+n_{\text{BE}}\left(\epsilon_{k}\right)\right]$,
$L_{1}\left(\epsilon_{1}\right)=2\alpha\eta\left(\Ket{E_{0}}\Bra{E_{1}}+\Ket{E_{2}}\Bra{E_{3}}\right)$,
and $L_{2}\left(\epsilon_{2}\right)=2\alpha\delta\left(-\Ket{E_{0}}\Bra{E_{2}}+\Ket{E_{1}}\Bra{E_{3}}\right)$.
We considered the spectral density to be $J\left(\omega\right)=\kappa/\pi$.

\section{Coefficient of performance, cooling power and injected power \label{sec:COP}}

\setcounter{figure}{0}
\setcounter{equation}{0}
\global\long\def\thefigure{B\arabic{figure}}
\global\long\def\theequation{B\arabic{equation}}

Here we derive the expressions for the coefficient of performance
(COP), cooling power, and injected power in terms of the interaction
energy and COP lag. The COP $\epsilon$ is written as

\begin{equation}
\epsilon=\frac{-\left\langle Q_{c}^{R}\right\rangle }{\left\langle W_{\text{net}}^{S}\right\rangle }=\left(1+\frac{\Delta V_{c}^{SR}}{\left\langle Q_{c}^{S}\right\rangle }\right)\frac{\left\langle Q_{c}^{S}\right\rangle }{\left\langle W_{\text{net}}^{S}\right\rangle },
\end{equation}
where we used Eq.~(\ref{eq: energy engineered cold reservoir}).
The first factor is the $\gamma$ parameter we now we work with the
second factor, which only depends on the refrigerant quantities. From
the first law of thermodynamics applied to a closed cycle, $\langle W_{1}^{S}\rangle+\langle W_{3}^{S}\rangle=-\langle Q_{h}^{S}\rangle-\langle Q_{c}^{S}\rangle$,
we can rewrite the second factor as

\begin{equation}
\frac{\left\langle Q_{c}^{S}\right\rangle }{\left\langle W_{\text{net}}^{S}\right\rangle }=-\frac{\left\langle Q_{c}^{S}\right\rangle }{\left\langle Q_{h}^{S}\right\rangle +\left\langle Q_{c}^{S}\right\rangle }=-\left(1+\frac{\beta_{h}\left\langle Q_{h}^{S}\right\rangle }{\beta_{h}\left\langle Q_{c}^{S}\right\rangle }\right)^{-1}.\label{eq:old COP}
\end{equation}
We now find an expression for the two heat quantities.

Applying the identity $D\left(\rho_{t}||\rho_{t}^{\text{eq}}\right)=\beta\left[U\left(\rho_{t}\right)-F_{t}^{\text{eq}}\right]-S\left(\rho_{t}\right)$
to the states at the end of the first and second strokes, respectively,
we obtain
\begin{equation}
D\left(\rho_{\tau_{1}}||\rho_{\tau_{2}}^{\text{eq,}c}\right)=\beta_{c}\left[U\left(\rho_{\tau_{1}}\right)-F_{\tau_{1}}^{\text{eq,}c}\right]-S\left(\rho_{\tau_{1}}\right),
\end{equation}
and
\begin{equation}
D\left(\rho_{\tau_{2}}||\rho_{\tau_{2}}^{\text{eq,}c}\right)=\beta_{c}\left[U\left(\rho_{\tau_{2}}\right)-F_{\tau_{2}}^{\text{eq,}c}\right]-S\left(\rho_{\tau_{2}}\right).
\end{equation}
From these two equations we can write 
\begin{align}
\left[U\left(\rho_{\tau_{2}}\right)-U\left(\rho_{\tau_{1}}\right)\right]= & \beta_{c}^{-1}\left[D\left(\rho_{\tau_{2}}||\rho_{\tau_{2}}^{\text{eq,}c}\right)-D\left(\rho_{\tau_{1}}||\rho_{\tau_{2}}^{\text{eq,}c}\right)\right]\nonumber \\
 & +\beta_{c}^{-1}\left[S\left(\rho_{\tau_{2}}\right)-S\left(\rho_{\tau_{1}}\right)\right],
\end{align}
where we used that $F_{\tau_{2}}^{\text{eq,}c}=F_{\tau_{1}}^{\text{eq,}c}$
since $H_{\tau_{2}}=H_{\tau_{1}}$ and the inverse temperature $\beta_{c}$
is the same.

In a similar way,
\begin{align}
\left[U\left(\rho_{\tau_{4}}\right)-U\left(\rho_{\tau_{3}}\right)\right]= & \beta_{h}^{-1}\left[D\left(\rho_{\tau_{4}}||\rho_{\tau_{4}}^{\text{eq,}h}\right)-D\left(\rho_{\tau_{3}}||\rho_{\tau_{4}}^{\text{eq,}h}\right)\right]\nonumber \\
 & +\beta_{h}^{-1}\left[S\left(\rho_{\tau_{4}}\right)-S\left(\rho_{\tau_{3}}\right)\right],
\end{align}
where we used that $F_{\tau_{4}}^{\text{eq,}h}=F_{\tau_{3}}^{\text{eq,}h}$
since $H_{\tau_{4}}=H_{\tau_{3}}$ and the inverse temperature $\beta_{h}$
is the same. Then, the heat absorbed by the refrigerant during the
second and fourth strokes are, respectively, 
\begin{equation}
\left\langle Q_{c}^{S}\right\rangle =\beta_{c}^{-1}\left[D\left(\rho_{\tau_{2}}||\rho_{\tau_{2}}^{\text{eq,}c}\right)-D\left(\rho_{\tau_{1}}||\rho_{\tau_{2}}^{\text{eq,}c}\right)\right]+\beta_{c}^{-1}\Delta S_{2}\label{eq:heat cold}
\end{equation}
and
\begin{equation}
\left\langle Q_{h}^{S}\right\rangle =-\beta_{h}^{-1}D\left(\rho_{\tau_{3}}||\rho_{\tau_{4}}^{\text{eq,}h}\right)+\beta_{h}^{-1}\Delta S_{4},\label{eq:heat hot}
\end{equation}
where we used the assumption that the final state of the fourth stroke
is always the thermal state, hence $D\left(\rho_{\tau_{4}}^{\text{eq,}h}||\rho_{\tau_{4}}^{\text{eq,}h}\right)=0$.

Substituting Eq.~(\ref{eq:heat hot}) into Eq.~(\ref{eq:old COP})
we obtain
\begin{equation}
\epsilon=-\gamma\left(1+\frac{\Delta S_{4}-D\left(\rho_{\tau_{3}}||\rho_{\tau_{4}}^{\text{eq,}h}\right)}{\beta_{h}\left\langle Q_{c}^{S}\right\rangle }\right)^{-1}.\label{eq:almost there}
\end{equation}
For conservation of entropy in the cyclic cycle, $\Delta S_{2}+\Delta S_{4}=0$,
we use $\Delta S_{2}$ from Eq.~(\ref{eq:heat cold}) into Eq.~(\ref{eq:almost there}).
Substituting we obtain

\begin{equation}
\epsilon=\gamma\left(\frac{1}{\epsilon_{\text{Carnot}}}+\mathcal{L}\right)^{-1},
\end{equation}
where $\epsilon_{\text{Carnot}}=\left(\beta_{c}/\beta_{h}-1\right)^{-1}$
is the Carnot COP and

\begin{equation}
\mathcal{L}=\frac{D\left(\rho_{\tau_{1}}||\rho_{\tau_{1}}^{\text{eq,}c}\right)-D\left(\rho_{\tau_{2}}||\rho_{\tau_{2}}^{\text{eq,c}}\right)+D\left(\rho_{\tau_{3}}||\rho_{\tau_{3}}^{\text{eq,}h}\right)}{\beta_{h}\langle Q_{c}^{S}\rangle}\label{00}
\end{equation}
is the COP lag. Just rearranging the terms we get Eq.~(\ref{eq:COP}). 

In order to operate in the refrigerator regime the engineered cold
reservoir must release heat, hence $-\left\langle Q_{c}^{R}\right\rangle >0$.
The first law applied to the refrigerant and engineered cold reservoir
gives $\Delta U^{SR}=\left\langle Q_{c}^{S}\right\rangle +\left\langle Q_{c}^{R}\right\rangle +\Delta V_{c}^{SR}=0$.
From the sign constraint we get $-\left\langle Q_{c}^{R}\right\rangle =\left\langle Q_{c}^{S}\right\rangle +\Delta V_{c}^{SR}>0$.
Dividing by $\left\langle Q_{c}^{S}\right\rangle $ on both sides
of the inequality we get $\gamma>0$. Therefore, the positivity of
the parameter $\gamma$ is a necessary condition to operate the refrigerator.

\section{Reaching the Otto COP\label{sec:COP Otto}}

\setcounter{figure}{0}
\setcounter{equation}{0}
\global\long\def\thefigure{C\arabic{figure}}
\global\long\def\theequation{C\arabic{equation}}

In order to know when the Otto COP is reached it is more instructive
to write the COP in terms another lag, which contains the quasistatic
reference state in the divergences. The quasistatic state $\rho_{\tau_{1}}^{\text{qs,}h}$,
respectively $\rho_{\tau_{3}}^{\text{qs},c}$, is the state reached
if the driving of the first, respectively the third, stroke is performed
quasistatically (infinitely slowly). Taking these states as the reference
states for the divergence one can show the identities
\begin{equation}
D\left(\rho_{\tau_{1}}||\rho_{\tau_{1}}^{\text{qs,}h}\right)=\beta_{h}\frac{\omega_{0}}{\omega_{\tau_{1}}}U\left(\rho_{\tau_{1}}\right)-\beta_{h}F_{0}^{\text{eq,}h}-S\left(\rho_{\tau_{1}}\right)\label{eq:quasistatic stroke 1}
\end{equation}
 and 
\begin{equation}
D\left(\rho_{\tau_{3}}||\rho_{\tau_{3}}^{\text{qs,}c}\right)=\beta_{c}\frac{\omega_{\tau_{1}}}{\omega_{0}}U\left(\rho_{\tau_{3}}\right)-\beta_{c}F_{\tau_{3}}^{\text{eq,}c}-S\left(\rho_{\tau_{3}}\right).\label{eq:quasistatic stroke 3}
\end{equation}
We note that these two identities are derived assuming a qubit or
harmonic oscillator refrigerant~\citep{Camati2019}. Using $S\left(\rho_{\tau_{1}}\right)=S\left(\rho_{0}\right)=\beta_{h}U\left(\rho_{0}\right)-\beta_{h}F_{0}^{\text{eq,}h}$
into Eq.~(\ref{eq:quasistatic stroke 1}) we obtain
\begin{equation}
D\left(\rho_{\tau_{1}}||\rho_{\tau_{1}}^{\text{qs,}h}\right)=\beta_{h}\frac{\omega_{0}}{\omega_{\tau_{1}}}U\left(\rho_{\tau_{1}}\right)-\beta_{h}U\left(\rho_{0}\right).\label{eq:quasi 1}
\end{equation}
Similarly, using $S\left(\rho_{\tau_{3}}\right)=S\left(\rho_{\tau_{2}}\right)=\beta_{c}U\left(\rho_{\tau_{2}}\right)-\beta_{c}F_{\tau_{2}}^{\text{eq,}c}-D\left(\rho_{\tau_{2}}||\rho_{\tau_{2}}^{\text{eq,}c}\right)$
into Eq.~(\ref{eq:quasistatic stroke 3}) we obtain
\begin{equation}
D\left(\rho_{\tau_{3}}||\rho_{\tau_{3}}^{\text{qs,}c}\right)=\beta_{c}\frac{\omega_{\tau_{1}}}{\omega_{0}}U\left(\rho_{\tau_{3}}\right)+D\left(\rho_{\tau_{2}}||\rho_{\tau_{2}}^{\text{eq,}c}\right)-\beta_{c}U\left(\rho_{\tau_{2}}\right).\label{eq:quasi 2}
\end{equation}
Substituting $U\left(\rho_{0}\right)$ from Eq.~(\ref{eq:quasi 1})
and $U\left(\rho_{\tau_{3}}\right)$ from Eq.~(\ref{eq:quasi 2})
to write
\begin{equation}
\frac{\beta_{h}\left\langle Q_{h}^{S}\right\rangle }{\beta_{h}\left\langle Q_{c}^{S}\right\rangle }=-\frac{\omega_{0}}{\omega_{\tau_{1}}}-\frac{\mathcal{F}}{\beta_{h}\left\langle Q_{c}^{S}\right\rangle },\label{eq:58}
\end{equation}
where
\begin{equation}
\mathcal{F}=D\left(\rho_{\tau_{1}}||\rho_{\tau_{1}}^{\text{qs,}h}\right)+\frac{\beta_{h}\omega_{0}}{\beta_{c}\omega_{\tau_{1}}}\left[D\left(\rho_{\tau_{3}}||\rho_{\tau_{3}}^{\text{qs,}c}\right)-D\left(\rho_{\tau_{2}}||\rho_{\tau_{2}}^{\text{eq,}c}\right)\right].\label{friction}
\end{equation}
The ratio in Eq.~(\ref{eq:58}) appears in the expression for the
COP in Eq.~(\ref{eq:old COP}). Substituting Eq.~(\ref{eq:58})
into Eq.~(\ref{eq:old COP}) we finally obtain
\begin{equation}
\epsilon=\gamma\left(\frac{1}{\epsilon_{\text{Otto}}}+\frac{\mathcal{F}}{\beta_{h}\left\langle Q_{c}^{S}\right\rangle }\right)^{-1},
\end{equation}
where $\epsilon_{\text{Otto}}=\omega_{\tau_{1}}/\left(\omega_{0}-\omega_{\tau_{1}}\right)$
is the Otto COP. The COP can be equivalently written as 
\begin{equation}
\epsilon=\gamma\frac{\epsilon_{\text{Otto}}}{1+\epsilon_{\text{Otto}}\frac{\mathcal{F}}{\beta_{h}\left\langle Q_{c}^{S}\right\rangle }}.
\end{equation}

With this expression for the COP we can see how one can reach the
Otto COP. If there is no coherence in the energy basis being generated
in the first or third strokes, a condition we satisfy by imposing
that the driving Hamiltonian commutes at different times, then one
can show that $\mathcal{F}=0$~\citep{Camati2019}. The resulting
COP, in this case, becomes $\epsilon=\gamma\epsilon_{\text{Otto}}$,
which is the COP that is valid for our refrigerator. If the parameter
$\gamma=1$, which is the case when the coupling between the refrigerant
and the engineered cold reservoir is sufficiently small, the COP reaches
the Otto COP, even operating at finite times.

\end{document}